\documentclass[aps,prl,showpacs,twocolumn]{revtex4}
\usepackage{graphicx}
\usepackage{epsfig}
\begin{document}
\title{ Parity-Affected Persistent Currents in
Superconducting Nanorings}
\author{Sergei V. Sharov$^{1,3}$}
\author{ Andrei D. Zaikin$^{2,4}$}
\affiliation{$^{1}$Institut f\"ur Theoretische Festk\"orperphysik,
Universit\"at Karlsruhe, 76128 Karlsruhe, Germany\\
$^{2}$Forschungszentrum Karlsruhe, Institut f\"ur Nanotechnologie,
76021 Karlsruhe, Germany\\
$^{{3}}$ Institute for Physics of Microstructures, Russian Academy of
Sciences, 603950 Nizhny Novgorod,
 Russia\\
$^{4}$I.E.Tamm Department of Theoretical Physics, P.N.Lebedev
Physics Institute, 119991 Moscow, Russia}

\date{\today}

\begin{abstract}
We argue that at sufficiently low temperatures $T$ superconducting
parity effect may strongly influence equilibrium persistent
currents (PC) in isolated superconducting nanorings containing a
weak link with few conducting modes. An odd electron, being added
to the ring, occupies the lowest available Andreev state and
produces a countercurrent circulating inside the ring. For a
single channel quantum point contact at $T=0$ this countercurrent
exactly compensates the supercurrent $I_e$ produced by all other
electrons and, hence, yields complete {\it blocking} of PC for any
value of the external magnetic flux. In superconducting nanorings
with embedded normal metal the odd electron countercurrent can
``overcompensate'' $I_e$ and a novel ``$\pi /N$-junction'' state
occurs in the system. Changing the electron parity number from
even to odd results in {\it spontaneous} supercurrent in the
ground state of such rings without any externally applied magnetic
flux.
\end{abstract}

\maketitle


Thermodynamic properties of isolated superconducting systems are
sensitive to the parity of the total number of electrons
\cite{AN92,Tuo93} even though this number ${\cal N}$ is
macroscopically large. This parity effect is a fundamental
property of a superconducting ground state described by the
condensate of Cooper pairs. The number of electrons in the
condensate is necessarily even, hence, for odd ${\cal N}$ at least
one electron remains unpaired having an extra energy equal to the
superconducting energy gap $\Delta$. This effect makes
thermodynamic properties of the ground states with even and odd
${\cal N}$ differ. Clear evidence for such parity effect was
demonstrated experimentally in small superconducting islands
\cite{Tuo93,Laf93}.

Can the supercurrent be affected by this parity effect? In many
structures the answer to this question is negative because of the
fundamental uncertainty relation $\delta {\cal N} \delta \varphi
\gtrsim 1$. Should the electron number ${\cal N}$ be fixed,
fluctuations of the superconducting phase $\varphi$ become large
disrupting the supercurrent. On the other hand, in transport
experiments with fluctuations of $\varphi$ being suppressed the
parity effect cannot be observed because of large fluctuations of
${\cal N}$.

Nonetheless, the parity effect {\it can}
coexist with non-vanishing supercurrent. The way out is to
consider systems supporting circular persistent currents (PC),
such as, e.g., isolated superconducting rings pierced by the
magnetic flux $\Phi$. In accordance with the number-phase
uncertainty relation the global superconducting phase of the ring
fluctuates strongly in this case, however these fluctuations are
decoupled from the supercurrent and therefore cannot influence the latter.
The aim of this
paper is to demonstrate that the parity effect may substantially
modify PC in superconducting nanorings, in particular if the total
number of electrons is odd.

Consider, for instance, a single mode quantum point contact (QPC)
embedded in a superconducting ring. Provided the number of
electrons in the system is not fixed the Josephson current across
this contact is given by the well known expression \cite{KO}
\begin{equation}
I(\varphi )=-\frac{2e}{\hbar}\frac{\partial \varepsilon (\varphi)}
{\partial \varphi} \tanh\frac{\varepsilon (\varphi)}{2T},
\label{josephson}
\end{equation}
where $\varphi$ is the phase difference across QPC, $\varepsilon
(\varphi)=\Delta \sqrt{1-{\cal T}\sin^2(\varphi/2)}$ and ${\cal
T}$ is the contact transmission. This result has a very
transparent physical interpretation being related \cite{Furusaki}
to contributions of discrete Andreev energy states $E_{\pm}=\pm
\epsilon (\varphi  )$ inside QPC via
\begin{equation}
I(\varphi )=\frac{2e}{\hbar}\left[\frac{\partial E_-}{\partial
\varphi} f_-(E_-) +\frac{\partial E_+}{\partial \varphi}
f_+(E_+)\right]. \label{jos2}
\end{equation}
Using the Fermi filling factors for these states $f_\pm (E_\pm
)=[1+\exp (\pm \epsilon (\varphi )/T)]^{-1}$ one arrives at Eq.
(\ref{josephson}).
\begin{figure}[b]
\vskip  -0.4 truecm
\centerline{\epsfig{file=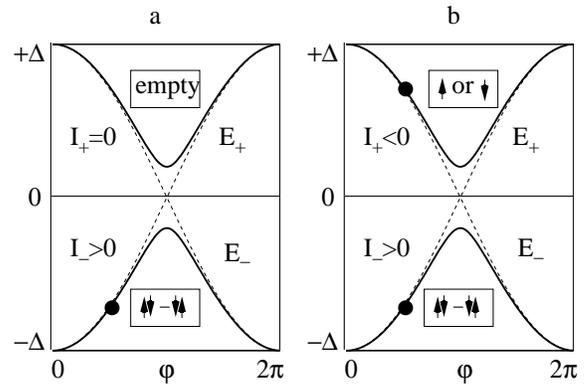,width=75mm}}
\caption{Andreev levels inside QPC and their occupation at
$T=0$ for even (a) and odd (b) ensembles.}
\end{figure}

Let us now fix the number of electrons inside the ring and
consider the limit $T \to 0$. For even ${\cal N}$ all electrons
are paired occupying available states with energies below the
Fermi level (see Fig. 1a). In this case one has $f_-(E_- )=1$,
$f_+ (E_+ )=0$, the current is entirely determined by the
contribution of the quasiparticle state $E_-$ and Eq. (\ref{jos2})
yields the same result as one for the grand canonical ensemble. By
contrast, in the case of odd ${\cal N}$ one electron always
remains unpaired and occupies the lowest available energy state --
in our case $E_+$ -- above the Fermi level. Hence, for odd ${\cal
N}$ one has $f_\pm (E_\pm )=1$ (Fig. 1b), the contributions of the
two Andreev states in Eq. (\ref{jos2}) exactly cancel each other,
and the current across QPC remains zero for any $\varphi$ or the
magnetic flux $\Phi$. Thus, we predict {\it a novel mesoscopic
effect} -- parity-induced blocking of PC in superconducting
nanorings with embedded QPC \cite{FNJ}.

It is important to emphasize that the above physical picture is
based on the assumption that the supercurrent is determined {\it
only} by the contributions of two discrete Andreev levels $E_{\pm}
(\varphi)$. This is the case provided (a) quasiparticle states of
the continuum do not contribute to the supercurrent and (b) there
are no more discrete Andreev levels inside the junction. Both
these conditions are met only for symmetric and extremely short
QPC. In a general case of $SNS$ junctions with a non-vanishing
thickness of the normal layer $d \neq 0$ these conditions are
violated, and no exact compensation of the supercurrent by the odd
electron countercurrent is anymore possible.

Further analysis reveals even richer physics in the latter case.
In particular, below we will demonstrate that at low $T$ and for
odd number of electrons a novel $\pi /N$-junction state should
occur in such $SNS$ rings. This effect in turn leads to {\it
spontaneous} supercurrent flowing in the ring in its ground state.
Under certain conditions the magnitude of this current can be as
high as $I_{sp} \sim e\Delta /\hbar$ reaching the values up to $10
\div 100$ nA for generic BCS superconductors. Spontaneous currents
of this magnitude can reliably be detected in modern experiments.

{\it Parity projection formalism}. In order to systematically
investigate an interplay between the parity effect and PC in
superconducting nanorings we will employ the parity projection
formalism \cite{JSA94,GZ94,AN94}.

The grand canonical partition function ${\cal Z}(T,\mu )= {\rm Tr
} e^{-\beta({\cal H}-\mu {\cal N})}$ is linked to the canonical
one $Z(T,{\cal N})$ as
\begin{equation}
\label{6}{\cal Z}(T,\mu )=\sum\limits_{{\cal N}=0}^\infty Z(T,{\cal N})\exp
\biggl({\frac{%
\mu {\cal N}}T}\biggr).
\end{equation}
Here and below ${\cal H}$ is the system Hamiltonian
and $\beta \equiv 1/T$. Inverting
this relation and defining the canonical partition functions
$Z_{e}$ and $Z_{o}$ respectively for even (${\cal N}\equiv {\cal
N}_{e}$) and odd (${\cal N} \equiv {\cal N}_{o}$) ensembles, one
gets
\begin{equation}
\label{8}Z_{e/o}(T)={\frac 1{2\pi }}\int\limits_{-\pi }^\pi due^{-i{\cal N}_
{e/o}u}%
{\cal Z}_{e/o}(T,iTu),
\label{invrel}
\end{equation}
where
$$
{\cal Z}_{e/o}(T,\mu )={1\over 2} {\rm Tr }\left\{\big[1\pm (-1)^{\cal N}\big]
e^{-\beta({\cal H}-\mu {\cal N})}\right\}
$$
\begin{equation}
\label{11}={\frac 12}\bigl({\cal Z}(T,\mu )\pm {\cal Z}(T,\mu +i\pi T)\bigr)%
\label{pm}
\end{equation}
are the parity projected grand canonical partition functions. For
${\cal N} \gg 1$ it is sufficient to evaluate the integral in
(\ref{invrel}) within the saddle point approximation
\begin{equation}
\label{14}Z_{e/o}(T)\sim e^{-\beta (\Omega _{e/o}-\mu _{e/o}{\cal
N}_{e/o})},
\end{equation}
where ${\Omega }_{e/o} =-T\ln {\cal Z}_{e/o}(T,\mu )$ are the
parity projected thermodynamic potentials,
\begin{eqnarray}
{\Omega }_{e/o}={\Omega }_{f} - T \ln \bigg[ \frac{1}{2} \Big( 1
\pm e^{- \beta ({\Omega_{b}} - {\Omega_{f}})} \Big) \bigg]
\label{Ome/o}
\end{eqnarray}
and ${\Omega}_{f/b}= - T \ln \left[ {\rm Tr }\left\{(\pm1)^{\cal
N} e^{-\beta({\cal H}-\mu {\cal N})}\right\} \right]$.
``Chemical potentials'' $\mu _{e/o}$ are
defined by the saddle point condition
${\cal N}_{e/o}=-\partial \Omega _{e/o}(T,\mu_{e/o})/\partial \mu_{e/o}$.

The main advantage of the above formalism is that it allows to
express the canonical partition functions and thermodynamical
potentials in terms of the parity projected grand canonical ones
thereby enormously simplifying the whole calculation. We further
note that $\Omega_{f}$ is just the standard grand canonical
thermodynamic potential and $\Omega_{b}$ represents the
corresponding potential linked to the partition function ${\cal
Z}(T,\mu +i\pi T)$. It is easy to see \cite{GZ94} that in order to
recover this function one can evaluate the true grand canonical
partition function ${\cal Z}(T,\mu )$, express the result as a sum
over the Fermi Matsubara frequencies $\omega_f =2\pi T(l+1/2)$ and
then substitute the Bose Matsubara frequencies  $\omega_b =2\pi
Tl$ ($l=0,\pm 1,...$) instead of $\omega_f$. This procedure
automatically yields the correct expression for ${\cal Z}(T,\mu
+i\pi T)$ and, hence, for $\Omega_{b}$.

Having found the thermodynamic potentials for the even and odd
ensembles one can easily determine the equilibrium current $I_{e/o}$.
Consider, as before, isolated superconducting rings pierced by the
magnetic flux $\Phi$. Making use of the above expressions one
finds PC circulating inside the ring:
\begin{equation}
I_{e/o}=I_{f}\pm \frac{I_{b}-I_{f}}{ e^{\beta ({\Omega_{b}} -
{\Omega_{f}})} \pm 1}, \label{Ie/o}
\end{equation}
where the upper/lower sign corresponds to the even/odd ensemble
and we have defined
$$
I_{e/o}= -c \left( \frac{\partial \Omega_{e/o}}{\partial \Phi}
\right) _{\mu(\Phi)},\;\;\; I_{f/b}= -c \left( \frac{\partial
\Omega_{f/b}}{\partial \Phi} \right) _{\mu(\Phi)}.
$$
{\it Homogeneous superconducting rings.} Turning to concrete
calculations we first treat homogeneous nanorings with
cross section $s$ and perimeter $L=2\pi R$. Throughout this
paper rings will be assumed sufficiently thin, $\sqrt{s} <
\lambda_L$, where $\lambda_L$ is the London penetration length.
Superconducting properties of such rings will be described within
the (parity projected) mean field BCS theory. At low temperatures
this description is justified provided quantum fluctuations of the
order parameter, quantum phase slips (QPS) \cite{QPSth}, can be
neglected. This requirement can be fulfilled only provided the
total number of conducting channels in the ring remains large
\cite{FN0} $ N_{\rm r} \sim p_F^2s \gg 1$. In addition, the
perimeter $L$ should not be too large in order to disregard
the QPS-induced reduction of PC \cite{MLG}. Finally,
we will neglect the difference between the mean field values of
the BCS order parameter for the even and odd ensembles
\cite{JSA94,GZ94}. This is legitimate provided the ring volume
is large enough, ${\cal V}=Ls \gg 1/\nu\Delta$, where $\nu$ is the
density of states at the Fermi level and $\Delta$ is the BCS order
parameter for a bulk superconductor at $T=0$.

The task at hand is now to evaluate the thermodynamic potentials
$\Omega_{f/b}$. Within the mean field approach these quantities
can be expressed in terms of the excitation energies
$\varepsilon_{k}$ and the order parameter
$\Delta(\mbox{\boldmath$r$})$. One finds \cite{GZ94}
\begin{eqnarray}
\Omega_{f}=  \tilde \Omega &-&2T \sum_{k} \ln\left(2 \cosh
\frac{\varepsilon_{k}}{2T} \right),
\label{Omf}\\
\Omega_{b}= \tilde \Omega &-&2T \sum_{k} \ln\left(2 \sinh
\frac{\varepsilon_{k}}{2T} \right), \label{Omb}
\end{eqnarray}
where $\tilde \Omega  = \int d^3 \mbox{\boldmath$r$}
|\Delta(\mbox{\boldmath$r$})|^2 / g +{\rm Tr}\{\hat \xi\}, $ $g$
is the BCS coupling constant and  $\hat \xi$ is the
single-particle energy operator:
\begin{equation}
\hat \xi=
\frac{1}{2m}{\left(-i\hbar\frac{\partial}{\partial\mbox{\boldmath$r$}}-
\frac{e}{c}\mbox{\boldmath$A$}(\mbox{\boldmath$r$})\right)}^2
+U(\mbox{\boldmath$r$})-\mu ,
\end{equation}
\mbox{\boldmath$A$}(\mbox{\boldmath$r$}) is the vector potential
and $U(\mbox{\boldmath$r$})$ describes the potential profile due
to disorder and interfaces.

The excitation spectrum $\varepsilon_{k}$ has the form
\begin{equation}
\varepsilon_{k}=\varepsilon(\mbox{\boldmath$p$})=\mbox{\boldmath$p$}
\mbox{\boldmath$v$}_{S}+
\sqrt{\xi^{2}+ \Delta^{2}},
\label{exc}
\end{equation}
where $\mbox{\boldmath$p$}$ is a quasiparticle momentum, $\xi
=(p^{2}-\tilde {\mu})/2m$, $\tilde
{\mu}=\mu(\Phi)-{m{v_{S}}^2}/2$ and
\begin{equation}
v_{S}= \frac{\hbar}{2mR}{\rm
min}_n\left(n-\frac{\Phi}{\Phi_{0}}\right)
\label{st}
\end{equation}
is the superconducting velocity. Both
$\varepsilon(\mbox{\boldmath$p$})$ and $v_S$  are periodic
functions of the flux $\Phi$ with the period equal to the
superconducting flux quantum $\Phi_{0}=hc/2e$.

Consider the most interesting case $T \ll
\hbar v_{F}/L$ \cite{FN6}. Making use of the
above expressions one obtains
\begin{equation}
I_{e}=ev_{S}\varrho_{e} s,\;\;\;\;
I_{o}=ev_{S}\varrho_{o}s-e\frac{v_{F}}{L}\mbox{sgn} v_{S},
\label{evenI}
\end{equation}
where $\varrho_{e/o}={\cal N}_{e/o}/{\cal V}$. The first Eq.
(\ref{evenI}) -- together with (\ref{st}) -- coincides with that
for the grand canonical ensemble. In contrast, for odd ensembles
(second Eq. (\ref{evenI})) there exists an additional
flux-dependent term which, however, remains rather small in
multichannel rings \cite{FN}. Estimating the main contribution to
$I_{e/o}$ as $I\sim ev_{F}N_{\rm r}/L$, one gets $(I_{e}-I_{o})/I
\sim 1/N_{\rm r} \ll 1$.
\begin{figure}[t]
\centerline{\epsfig{file=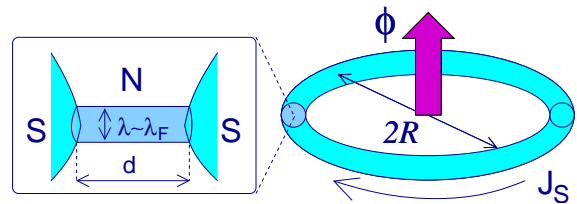,width=75mm}}
\caption{Superconducting ring with embedded $SNS$ junction of length
$d$.}
\end{figure}

Now we proceed to the situations in which parity effect gains more
importance and leads to new physical effects. Namely, we consider
superconducting rings with $N_{\rm r}\gg 1$ interrupted by a weak
link with only few conducting channels $N$. In such systems the
mean field BCS description remains applicable and, on the other
hand, the parity effect can be large due to the condition $N \sim
1$. Quantum point contacts (QPC) or, more generally, $SNS$
junctions (of length $d$) can be used for practical realization of
such weak links. The corresponding structure is depicted in Fig.
2.

{\it Superconducting rings with QPC}. As a first example we consider a ring
interrupted by QPC. The optimal value of the phase difference $\varphi$
across QPC is fixed by the equation $ \partial \Omega_{e/o}/\partial \varphi
=0$, which yields the current conservation condition
$I^{(c)}=I^{(r)}$. Here $I^{(c)} \sim
2e{\cal T}\Delta /\hbar$ is the current through the contact
and $I^{(r)} \simeq (ev_{F}N_{\rm r}/L)(\varphi-2\pi \Phi /\Phi_{0})$ is the
supercurrent inside the ring. From the current conservation one easily finds
\cite{FN77}
\begin{eqnarray}
\label{L<}\varphi&\simeq &2\pi \Phi /\Phi_{0},\;\;\;\mbox{  if  } \;\; L\ll L^
{*},\\
\varphi&\simeq &2\pi n, \;\;\;\;\;\;\;\;\; \mbox{     if  }\;\; L\gg
L^{*}, \label{L>}
\end{eqnarray}
where $L^{*}=\xi_{0}N_{\rm r}/{\cal T} \gg \xi_0\sim \hbar v_F/\Delta$. In a
more general case of QPC with $N$ conducting channels in the expression
for $L^{*}$ one should set ${\cal T} \to \sum_i^N{\cal T}_i$.

In the limit (\ref{L>}) one recovers the results identical to ones
for homogeneous rings. Below we will concentrate on the opposite
limit $L \ll L^{*}$ and also assume $N \ll N_{\rm r}$. Due to Eq.
(\ref{L<}) in this case the dependence $I_{e/o}(\Phi )$ is fully
determined by the current-phase relation for QPC which can be
found by means of Eq. (\ref{Ie/o}). Expressing $I_{f/b}$ via $\varphi$
it is convenient to employ the general formula \cite{GZ}
\begin{equation}
I_{f/b}=\frac{2e}{\hbar}\sum_{i=1}^N T\sum_{\omega_{f/b}}
\frac{\sin\varphi}{\cos\varphi + W_i(\omega_{f/b})}. \label{fbfb}
\end{equation}
For QPC one has $W_i(\omega )= (2/{\cal
T}_i)(1+\hbar^2\omega^2/\Delta^2)-1$. Substituting this function into
(\ref{fbfb}) and summing over $\omega_f$ one rederives the standard
result \cite{KO}, while the same summation over $\omega_b$ yields
\begin{equation}
I_{b}=-\frac{2e}{\hbar}\sum_{i=1}^N \frac{\partial
\varepsilon_{i}(\varphi)} {\partial \varphi}
\coth\frac{\varepsilon_{i}(\varphi)}{2T}, \label{bose}
\end{equation}
where $\varepsilon_{i}(\varphi)=\Delta \sqrt{1-{\cal
T}_i\sin^2(\varphi/2)}$. Finally, the difference
$\Omega_b-\Omega_f\equiv \Omega_{bf}$ is evaluated as a sum of the
ring ($\Omega_{bf}^{(r)}$) and QPC ($\Omega_{bf}^{(c)}$)
contributions. The latter is found by integrating $I_{f/b}(\varphi
)$ over the phase $\varphi$ with the result
\begin{equation}
\Omega_{bf}^{(c)}=2T\sum_{i=1}^N \ln
\coth\left(\frac{\varepsilon_{i}(\varphi)}{2T}\right),
\end{equation}
while the former is defined by the standard expression
$$
\beta\Omega_{bf}^{(r)}=2{\cal V}\int \frac{d^3\mbox{\boldmath$p$}}
{(2\pi \hbar)^3}\ln\left(\coth
\frac{\varepsilon(\mbox{\boldmath$p$})}{2T} \right)\simeq \nu
{\cal V}\sqrt{\Delta T}e^{-\frac{\Delta }{T}}.
$$
Combining all these results with Eq. (\ref{Ie/o}) we get
$$
I_{e/o}=-\frac{2e}{\hbar}\sum_{i=1}^N \frac{\partial
\varepsilon_{i}(\varphi)} {\partial \varphi}
\tanh\frac{\varepsilon_{i}(\varphi)}{2T}
$$
\begin{equation}
\times \left[1\pm \frac{(\coth\frac
{\varepsilon_{i}(\varphi)}{2T})^{2}-1}{e^{\beta\Omega_{bf}^{(r)}}\prod\limits_
{j=1}^N
(\coth\frac{\varepsilon_{j}(\varphi)}{2T}) ^{2} \pm 1}\right].
\label{result}
\end{equation}
\begin{figure}[b]
\vspace{0.3cm}
\centerline{\epsfig{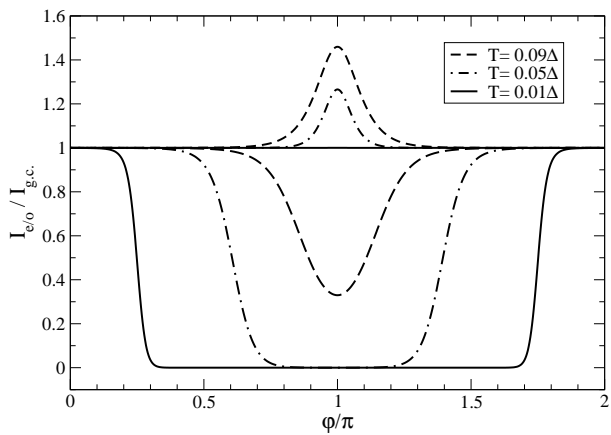}}
\caption{The ratio between canonical and grand canonical values of
PC $I_{e/o}/I_{g.c.}$ (represented by the term in the square
brackets in Eq. (\ref{result})) versus $\varphi$ in a single mode
QPC at different temperatures for even (three upper curves) and
odd (three lower curves) ensembles. Here we have chosen
$T^*_S=0.1\Delta$ and ${\cal T}=0.99$.}
\end{figure}
The term in the square brackets accounts for the parity effect in
our system. For $N=1$ and at $T=0$ this term reduces to unity for
even ensembles and to zero for odd ones, thus confirming our
intuitive picture of parity-induced blocking of PC discussed
above. For $T >0$ Eq. (\ref{result}) demonstrates that both for
even and especially for odd ${\cal N}$ the current-phase relation
for QPC may substantially deviate from that derived for the grand
canonical ensemble \cite{KO}, see Fig. 3. For even ensembles the
supercurrent {\it increases} above its grand canonical value. This
effect is mainly pronounced for phases $\varphi$ not very far from
$\varphi =\pi$ and -- at sufficiently low $T$ -- it becomes
progressively more important with increasing temperature. On the
contrary, for odd ensembles the supercurrent is always suppressed
below its grand canonical value. This suppression is gradually
lifted with increasing temperature, though at phases $\varphi$ in
the vicinity of the point $\varphi =\pi$ blocking of PC may
persist up to sufficiently high $T$. Eq. (\ref{result}) also shows
that in QPC with several conducting channels and at $T \to 0$ the
current through the most transparent channel will be blocked by
the odd electron. Hence, though blocking of PC remains incomplete
in this case, it may nevertheless be important also for QPC with
$N>1$.

{\it SNS rings.} Finally, we turn to superconducting rings
containing a piece of a normal metal with non-zero thickness $d$.
In contrast to the situation of QPC considered above, the
Josephson current in $SNS$ structures cannot anymore be attributed
only to the discrete Andreev states inside a weak link, and an
additional contribution from the states in the continuum should
also be taken into account. Furthermore, for any non-zero $d$
there are always more than two discrete Andreev levels in the
system. Accordingly, significant modifications in the physical
picture of the parity effect in $SNS$ rings can be expected.
\begin{figure}[b]
\centerline{\epsfig{file=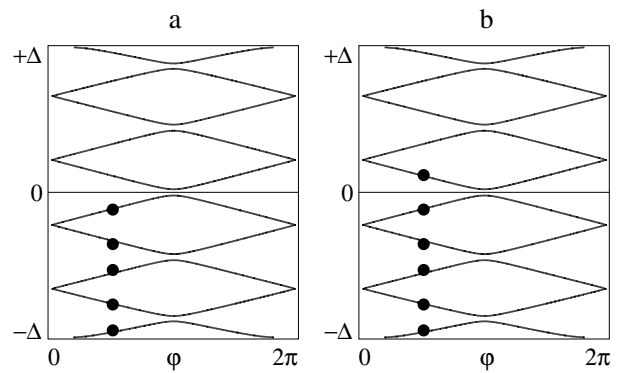,width=80mm}}
\caption{Andreev levels in a single mode $SNS$ junction with
$d=6\hbar v_F/\Delta$ and their occupation at $T=0$ for even (a)
and odd (b) ensembles.}
\end{figure}

The key difference can be understood already by comparing the
typical structure of discrete Andreev levels in $SNS$ junctions
(Fig. 4) with that of QPC (Fig. 1).  As before, in the limit $T
\to 0$ all states below (above) the Fermi level are occupied
(empty) provided the total number of electrons in the system is
even (Fig. 4a). If, on the other hand, this number is odd the
lowest Andreev state above the Fermi energy is occupied as well
(Fig. 4b) thus providing an additional contribution to the
Josephson current. This contribution, however, cancels only that
of a symmetric Andreev level below the Fermi energy, while the
contributions of all other occupied Andreev levels and of the
continuum states remain uncompensated. Hence, unlike in the QPC
limit displayed in Fig. 1, in $SNS$ rings one should not anymore
expect the effect of PC blocking by the odd electron, but rather
some other non-trivial features of the parity effect.
\begin{figure}[b]
\centerline{\epsfig{file=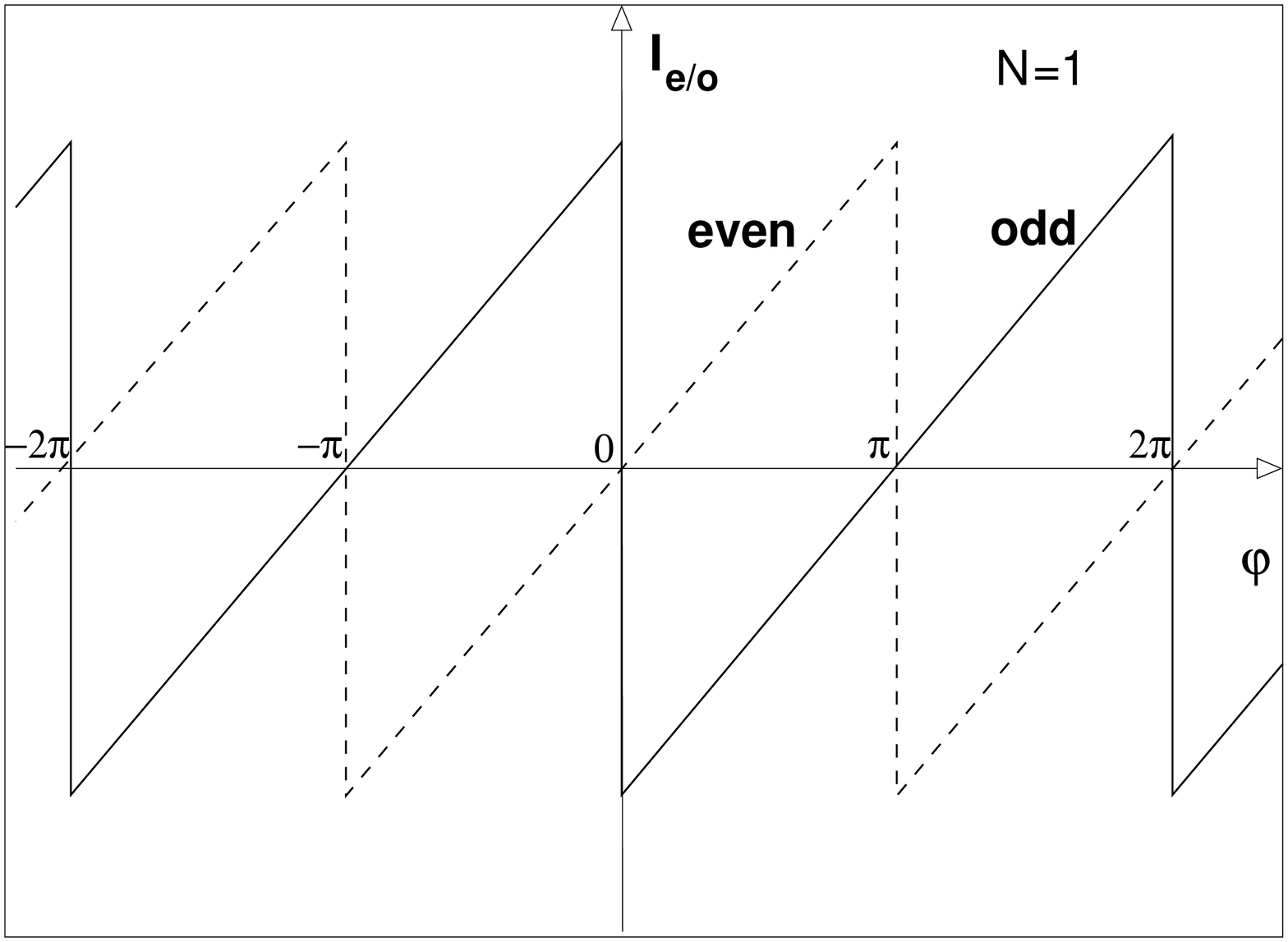,width=75mm}}
\caption{The zero temperature current-phase dependence (\ref{SNS})
for $SNS$ rings with $N=1$: $I_{e}(\varphi)$ (dashed) line and
$I_{o}(\varphi)$ (solid) line.}
\end{figure}

In order to construct a quantitative description of the parity
effect in $SNS$ rings we will again make use of the parity
projection technique formulated above. Here we only restrict
our attention to transparent $SNS$ junctions where the effects
under consideration are most pronounced. In this case the function
$W_i (\omega )\equiv W(\omega )$ is the same for all channels. It
reads
$$
W (\omega )= \left(\frac{2\hbar^2\omega^2}{\Delta^2}+1\right)
\cosh \left(\frac{2\omega d}{v_F}\right)
$$
\begin{equation}
+
\frac{2\hbar\omega}{\Delta}\sqrt{1+\frac{\hbar^2\omega^2}{\Delta^2}}
\sinh \left(\frac{2\omega d}{v_F}\right).
\end{equation}
Substituting this function into (\ref{fbfb}) and repeating the
whole calculation as above, we arrive at the final result which
cannot in general be represented in a tractable analytic form.
Significant simplifications occur in the most interesting limit
\cite{FN3} $T \to 0$, in which case we obtain
\begin{eqnarray}
I_e=\frac{e\Delta N}{\hbar}\left(\sin\frac{\varphi}{2}-
\frac{2y\sin \varphi}{\pi}\ln \frac{1}{y}\right), \\
\label{SCSe}
I_o=I_e-\frac{e\Delta}{\hbar}\left(\sin \frac{\varphi}{2}
+y\mbox{sgn} \varphi \cos \varphi \right)
\label{SCSo}
\end{eqnarray}
for short $SNS$ junctions $y\equiv d\Delta /\hbar v_F \ll 1$ and
\begin{equation}
I_e=\frac{ev_FN}{\pi d}\varphi ,\;\;\;\;I_o=\frac{ev_FN}{\pi
d}\left(\varphi -\frac{\pi \mbox{sgn} \varphi}{N}\right)
\label{SNS}
\end{equation}
for long ones $d \gg \xi_0 \sim \hbar v_F/\Delta$. These results
apply for $-\pi <\varphi<\pi$ and should be $2\pi$-periodically
continued otherwise. The term containing $\ln (1/y)$ in Eq.
(\ref{SCSe}) for $I_e$ is written with the logarithmic accuracy
and is valid for $\varphi$ not too close to $\varphi =\pm\pi$.

Let us concentrate on the results (\ref{SNS}) applicable for
sufficiently long $SNS$ junctions. We observe that at $T=0$ the
current $I_e$ -- as in the QPC case -- coincides with that for the
grand canonical ensembles \cite{Kulik}, while for odd ${\cal N}$
the current-phase relation is shifted by the value $\pi /N$. This
shift has a simple interpretation being related to the odd
electron contribution $(2e/\hbar )\partial E_1/\partial \varphi$
from the lowest (above the Fermi level) Andreev state $E_1(\varphi
)$ inside the $SNS$ junction. As we have expected, this
contribution indeed does not compensate for the current from other
quasiparticle states. Rather it provides a possibility for a
parity-induced $\pi$-junction state \cite{Leva} in our system:
According to Eq. (\ref{SNS}) for single mode $SNS$ junctions the
``saw tooth'' current-phase relation will be shifted exactly by
$\pi$, see Fig. 5. More generally, we can talk about a novel $\pi
/N$-junction state, because in the odd case the minimum Josephson
energy (zero current) state is reached at $\varphi =\pm \pi /N$,
see Fig. 6. For any $N>1$ this is a twofold degenerate state
within the interval $-\pi <\varphi <\pi$ \cite{BGZ}. In the
particular case $N=2$ the current-phase relation $I_o(\varphi )$
turns $\pi$-periodic.
\begin{figure}[b]
\centerline{\epsfig{file=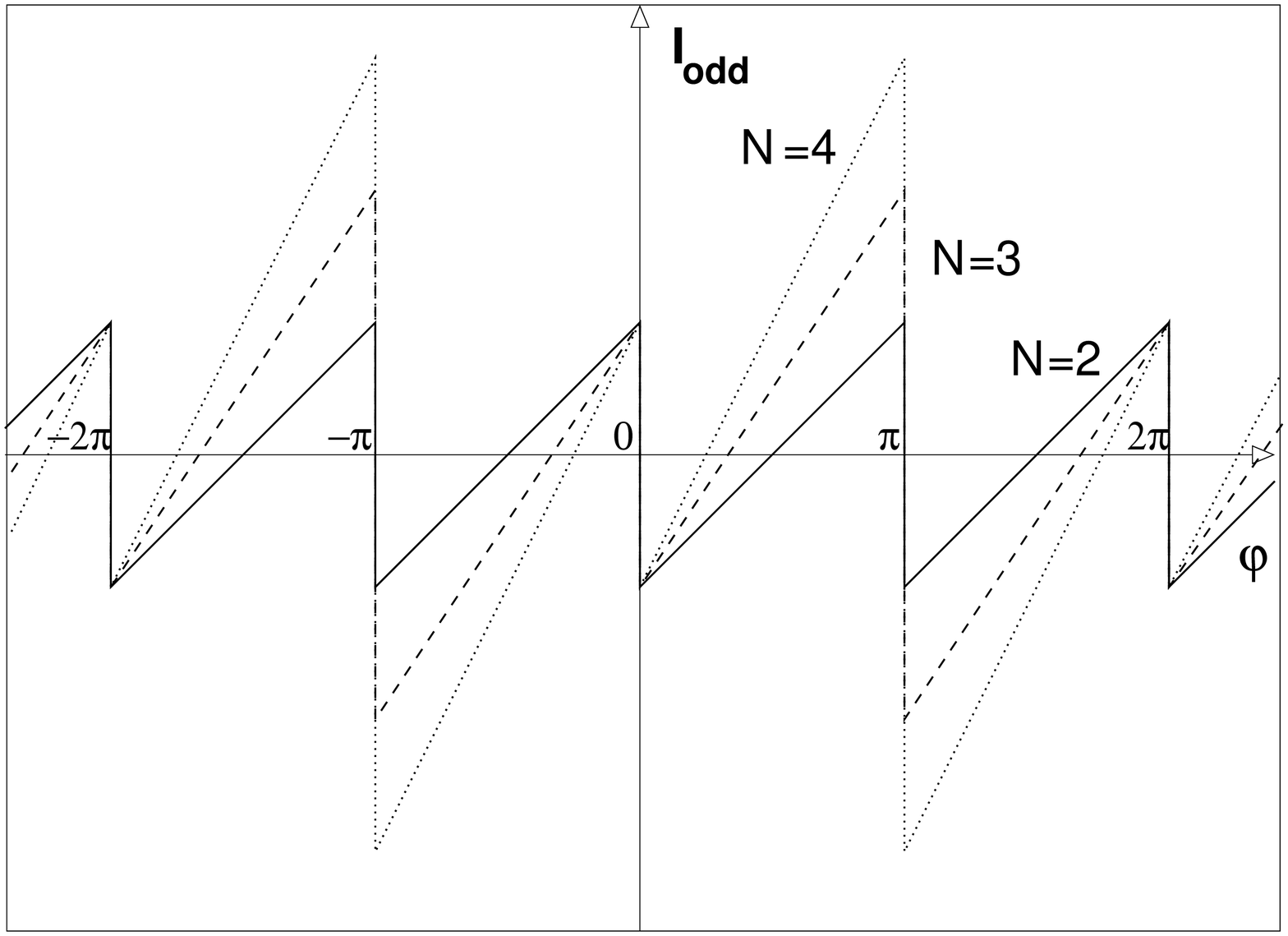,width=75mm}}
\caption{The same as in Fig. 5 only for the odd ensembles (second
Eq. (\ref{SNS})) and for $N=2, 3$ and $4$.}
\end{figure}

Let us recall that the $\pi$-junction state can be realized in
$SNS$ structures by driving the electron distribution function in
the contact area out of equilibrium \cite{Volkov,WSZ,Yip}. Here,
in contrast, the situation of a $\pi$- or $\pi /N$-junction is
achieved in thermodynamic equilibrium. Despite this drastic
difference, there also exists a certain physical similarity
between the effects discussed here and in Refs.
\onlinecite{Volkov,WSZ,Yip}: In both cases the electron
distribution function in the weak link deviates substantially from
the Fermi function. It is this deviation which is responsible for
the appearance of the $\pi$-junction state in both physical
situations.

It is also important to emphasize that the countercurrent produced
by the odd electron causes a jump on the current-phase dependence
at $\varphi=0$, and the direction of $I_o$ at sufficiently small
$\varphi$ is always opposite to that of $I_e$. This is a generic
feature of odd ensembles which persists for all non-zero values of
$y$. The current-phase relation $I_o(\varphi )$ for arbitrary
values of the parameter $y$ can be computed numerically. Examples
are presented in Fig. 7. One clearly observes the current jump at
$\varphi =0$. As it is obvious from Eq. (\ref{SCSo}), this feature
disappears only in the QPC limit $y \to 0$.

{\it BCS ground state with spontaneous current.} Perhaps the most
spectacular physical consequence of this current jump is the
presence of spontaneous supercurrents in the ground state of $SNS$
rings with odd number of electrons. Similarly to the case of
standard $\pi$-junctions \cite{Leva} such spontaneous
supercurrents  should flow even in the absence of an externally
applied magnetic flux. Unlike in Ref. \onlinecite{Leva}, however, here the
spontaneous current state occurs for {\it any} inductance of the
ring because of the non-sinusoidal dependence $I_o(\varphi )$.
\begin{figure}[t]
\vspace{0.5cm}
\centerline{\epsfig{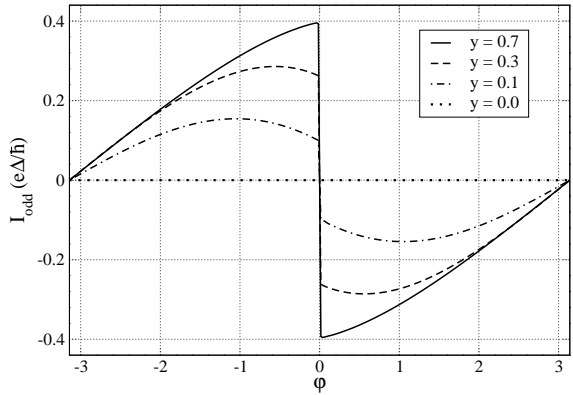}}
\caption{The zero temperature current-phase relation $I_o(\varphi
)$ ($-\pi <\varphi<\pi$)
 for $N=1$ and different values of the parameter
$y=d\Delta /\hbar v_F$.}
\end{figure}

Consider, for instance, the limit $d \gg \xi_0$. In the case of
odd number of electrons the ground state energy of an $SNS$ ring
can be written in a simple form
\begin{equation}
E=\frac{\Phi^2}{2c{\cal L}}+\frac{\pi\hbar
v_FN}{\Phi_0^2d}\left(\Phi -\frac{\Phi_0 \mbox{sgn}
\Phi}{2N}\right)^2,
\label{E}
\end{equation} where the first term is the magnetic energy of the ring
(${\cal L}$ is the ring inductance and $c$ is the speed of light)
while the second term represents the Josephson energy of the $SNS$
junction. Minimizing (\ref{E}) with respect to the flux $\Phi$ one
immediately concludes that the ground state of the ring is a
twofold degenerate state with a non-vanishing {\it spontaneous}
current
\begin{equation}
I=\pm \frac{ev_F}{d}\left[1 +\frac{2ev_FN}{d}\frac{{\cal
L}}{\Phi_0}\right]^{-1} \label{Isp}
\end{equation}
flowing either clockwise or counterclockwise. In the limit of
small inductances ${\cal L} \to 0$ this current does not vanish
and its amplitude just reduces to that of the odd electron current
at $\varphi \to 0$. Hence, at small ring inductances the magnitude
$I_{sp}$ of this spontaneous current can easily be obtained from
Eqs. (\ref{SCSo}-\ref{SNS}):
\begin{eqnarray}
I_{sp}= e\Delta^2d/\hbar^2 v_F,\;\;\;\; &&\text{if}\;\;d \ll
\xi_0,
\label{sp1}\\
I_{sp}= ev_F/\pi d,\;\;\;\; &&\text{if}\;\;d \gg \xi_0.\label{sp2}
\end{eqnarray}
For intermediate values of the parameter $y$ the amplitude of the
current $I_{sp}$ can be evaluated numerically. The results are
displayed in Fig. 8. One observes that -- in agreement with Eq.
(\ref{sp1}) -- $I_{sp}$ increases linearly with $d$ at small $d$,
reaches its maximum value $I_{\rm max} \sim 0.4e\Delta/\hbar $ at
$d \sim \xi_0$ and then decreases with further increase of $d$
approaching the dependence (\ref{sp2}) in the limit of large $d$.
For generic BCS superconductors the magnitude of this maximum
current can be estimated as $I_{\rm max} \sim 10 \div 100$ nA.
These values might be considered as surprisingly large ones having
in mind that this current is associated with only one Andreev
electron state.
\begin{figure}[t]
\vspace{0.5cm}
\centerline{\epsfig{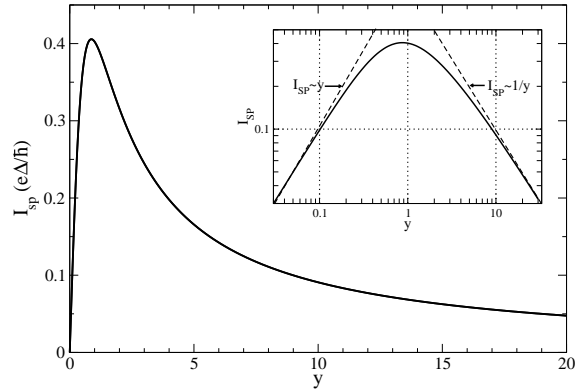}}
\caption{The  spontaneous current amplitude $I_{sp}$ as a function of
the parameter $y$ at $T=0$. In the inset, the same function is shown on
the $log-log$ scale. Dashed lines indicate the
asymptotic behavior of $I_{sp}(y)$ in the limits of small and large $y$.}
\end{figure}

In summary, new physical effects emerge from an interplay between
the electron parity number and persistent currents in
superconducting nanorings. Perhaps the most striking observation
is that the BCS ground state of such rings with the odd number of
electrons is the state with non-zero {\it spontaneous}
supercurrent. This and other novel features predicted here can be
directly tested in modern experiments.

This work is part of the
Kompetenznetz ``Funktionelle Nanostructuren'' supported by the Landestiftung
Baden-W\"urttemberg gGmbH and of the STReP project ``Ultra-1D'' supported by
the EU.



\begin{thebibliography}{99}
\bibitem{FN0} More detailed requirements can be formulated with the aid of
the results \cite{QPSth}. For generic wires QPS effects are small
provided the parameter  $\sqrt{s}$ exceeds $\sim 10$ nm
\cite{QPSth,QPSexp}.
\bibitem{FN77} Quantum fluctuations of $\varphi$ can easily be
suppressed, e.g. by choosing the ring inductance sufficiently
small. Accordingly, the number of Cooper pairs passing through the contact
should strongly fluctuate effectively suppressing electrostatic effects in the
contact area.
\bibitem{FN3} As it was already pointed out, the parity effects persist at sufficiently
low temperatures and disappear otherwise. For rings with
sufficiently short weak links the corresponding crossover
temperature is $T^*_S$ \cite{FN6}. In the case of long $SNS$
junctions $d \gg \xi_0$ and for not very large $N$ the estimate of
this crossover temperature is fully analogous and just amounts to
calculating the number of states available for a quasiparticle at
a given $T$. This estimate gives rise to the crossover temperature
$T^*_N \sim \hbar v_F/d\ln (NT^*_Nd/v_F)$ above which even-odd
effects in long $SNS$ junctions should be smeared out. In a
general case the crossover temperature is defined by either
$T^*_S$ or $T^*_N$ whatever is smaller.
\end{thebibliography}
\end{document}